\documentstyle[psfig]{l-aa-ps}

\def\mincir{\raise -2.truept\hbox{\rlap{\hbox{$\sim$}}\raise5.truept
\hbox{$<$}\ }}
\def\magcir{\raise -2.truept\hbox{\rlap{\hbox{$\sim$}}\raise5.truept
\hbox{$>$}\ }}

   \thesaurus{Section 1   % A&A Section 6: Form. struct. and evolut. of stars
              (02.18.5, 13.07.1)}  % gamma ray: bursts

\title{Optical and Near-Infrared Follow-up Observations of GRB980329}

\author{E. Palazzi$^1$ \and 
E. Pian$^1$ \and 
N. Masetti$^1$ \and  
L. Nicastro$^2$ \and  
P. Vreeswijk$^3$ \and  
T.J.  Galama$^3$ \and  
P. Groot$^3$ \and  
F. Frontera$^{1,4}$ \and  
M. Della Valle$^5$ \and  
C. Lidman$^6$ \and 
C. Kouveliotou$^7$ \and  
G. Pizzichini$^1$ \and 
J. van Paradijs$^{3,8}$ \and  
H. Pedersen$^{9}$ \and  
F. Mannucci$^{10}$ \and
M. Di Martino$^{11}$ \and
A.H. Diercks$^{12}$ \and
E.W. Deutsch$^{12}$ \and
L. Amati$^{13}$ \and 
S. Benetti$^{14}$ \and
A.J. Castro-Tirado$^{15,16}$ \and 
J. Clasen$^{17}$ \and 
E. Costa$^{13}$ \and  
D. Dal Fiume$^1$ \and  
R. Falomo$^{18}$ \and 
M. Feroci$^{13}$ \and  
J. Fynbo$^{19}$ \and 
J. Heise$^{20}$ \and
J.J.M. in 't Zand$^{20}$ \and
L. Piro$^{13}$ \and 
C. Robinson$^7$ \and 
M. Tornikoski$^{21}$ \and 
E. Valtaoja$^{22}$ \and
M.R. Zapatero-Osorio$^{23}$ \and
D. Lamb$^{24}$ \and
J. Quashnock$^{24}$ \and
D. Van den Berk$^{25}$}

\offprints{eliana@tesre.bo.cnr.it}

\institute{
$^1$ Istituto Tecnologie e Studio Radiazioni Extraterrestri, CNR, 
Via Gobetti 101, I-40129 Bologna, Italy\\
$^2$ Istituto di Fisica Cosmica con Applicazioni all'Informatica, CNR,
 Via U. La Malfa 153, I-90146 Palermo, Italy\\
$^3$ University of Amsterdam, Kruislaan 403, 1098 SJ Amsterdam, The 
Netherlands\\
$^4$ Dept. of Physics, University of Ferrara, Via Paradiso 11,
 I-44100 Ferrara, Italy\\
$^5$ Dept. of Astronomy, University of Padova, V. Dell'Osservatorio 5,
I-35122 Padova, Italy\\
$^6$ European Southern Observatory, Casilla 19000, Santiago, Chile\\
$^7$ Universities Space Research Association, Huntsville, AL, USA\\
$^8$ University of Alabama in Huntsville, 35899 AL, USA\\
$^9$ Copenhagen University Observatory, Juliane Maries Vej 30, DK 2100
Copenhagen, Denmark\\
$^{10}$ Centro per l'Astronomia Infrarossa e lo Studio del Mezzo Interstellare, 
CNR, Largo E. Fermi 5, I-50125 Arcetri, Italy\\
$^{11}$ Torino Astronomical Observatory, Pino Torinese, Italy\\
$^{12}$ Dept. of Astronomy, Box 351580, University of Washington,
Seattle, WA 98195 \\
$^{13}$ Istituto di Astrofisica Spaziale CNR, Via del Fosso del Cavaliere, 
I-00131 Roma, Italy\\
$^{14}$ Telescopio Nazionale Galileo, La Palma, Canary Islands, Spain\\
$^{15}$ Laboratorio de Astrof\'{\i}sica Espacial y F\'{\i}sica Fundamental 
(LAEFF-INTA), P.O. Box 50727, E-28080, Madrid, Spain\\
$^{16}$ Instituto de Astrof\'{\i}sica de Andaluc\'{\i}a (IAA-CSIC), P.O.
Box 03004, E-18080, Granada, Spain \\
$^{17}$ Nordic Optical Telescope, La Palma, Canary Island, Spain\\
$^{18}$ Astronomical Observatory of Padova, V. Dell'Osservatorio 5,
I-35122 Padova, Italy\\
$^{19}$ Institute of Physics and Astronomy, Aarhus, Denmark\\
$^{20}$ Space Research Organization in the Netherlands, Sorbonnelaan 2,
 3584 CA Utrecht, The Netherlands\\
$^{21}$ Mets\"ahovi Radio Observatory, Helsinki, Finland\\
$^{22}$ Tuorla Observatory, Turku, Finland\\
$^{23}$ Instituto de Astrof\'{\i}sica de Canarias, Via Lactea, La Laguna,
Tenerife, Spain\\
$^{24}$ Dept. of Astronomy \& Astrophysics, University of Chicago,
5640 South Ellis Av., Chicago, IL 60637, USA\\
$^{25}$ Mc Donald Observatory, University of Texas, RLM 15.308, Austin,
TX 78712-1083, USA}

\begin{document}

%   \date{Received June 4, 1998; accepted ....}

%\begin{document}
   \maketitle
   \markboth{E. Palazzi et al.: Optical and NIR observations of GRB980329}{}

   \begin{abstract}

We imaged the field of GRB980329 in the optical and in the near-infrared
starting 20 hours after the event, at the ESO NTT, at the NOT, and at the
TIRGO.  In the first night we detect an object of $R$ = 23.6 $\pm$ 0.2
within the BeppoSAX NFI error box at the same position as a transient VLA
source proposed as the radio afterglow of this GRB. The source faded by
1.6 $\pm$ 0.5 magnitudes in 2.1 days, similarly to the decays of previous
GRB optical afterglows.  This transient is likely the optical counterpart
of GRB980329. In the near-infrared we detect signal at 2--$\sigma$
significance, whose position is only marginally consistent with that of
the VLA source.  The spectrum of the transient bears the signatures of
substantial absorption within the GRB host galaxy.  The afterglow
energetics are interpreted as synchrotron radiation from an expanding
blast wave.

\end{abstract}

%
%________________________________________________________________

\section{Introduction}

The Gamma Ray Burst GRB980329 was detected on March 29.1559 UT with the
BeppoSAX Gamma Ray Burst Monitor (GRBM)  and Wide Field Cameras (WFC)  unit
number 2 (in 't Zand et al. 1998), and with BATSE (Briggs et al. 1998). The
burst, rapidly localized by the WFC with a 3${^\prime}$ radius accuracy, was
outstandingly bright in $\gamma$-rays (peak intensity of $\sim$6000 counts
s$^{-1}$ and fluence of $5.5 \times 10^{-5}$ erg cm$^{-2}$, 40-700 keV)  and
X-rays ($\sim$6 Crab, 2-26 keV).  Observations with the BeppoSAX Narrow
Field Instruments (NFI) started on March 29.449 UT revealed an unknown
fading X-ray source in the WFC error box which has been identified as the
GRB X-ray afterglow (in 't Zand et al. 1998).  Here we report on optical and
near-infrared imaging at the 3.5m ESO New Technology Telescope (NTT, La
Silla, Chile), at the 2.5m Nordic Optical Telescope (NOT, La Palma, Canary
Islands), at the ARC 3.5m telescope of the Apache Point Observatory (APO,
Arizona)  and at the 1.5m Gornergrat Infrared Telescope (TIRGO,
Switzerland), respectively. 

\begin{figure*}[t]
\centerline{\psfig{figure=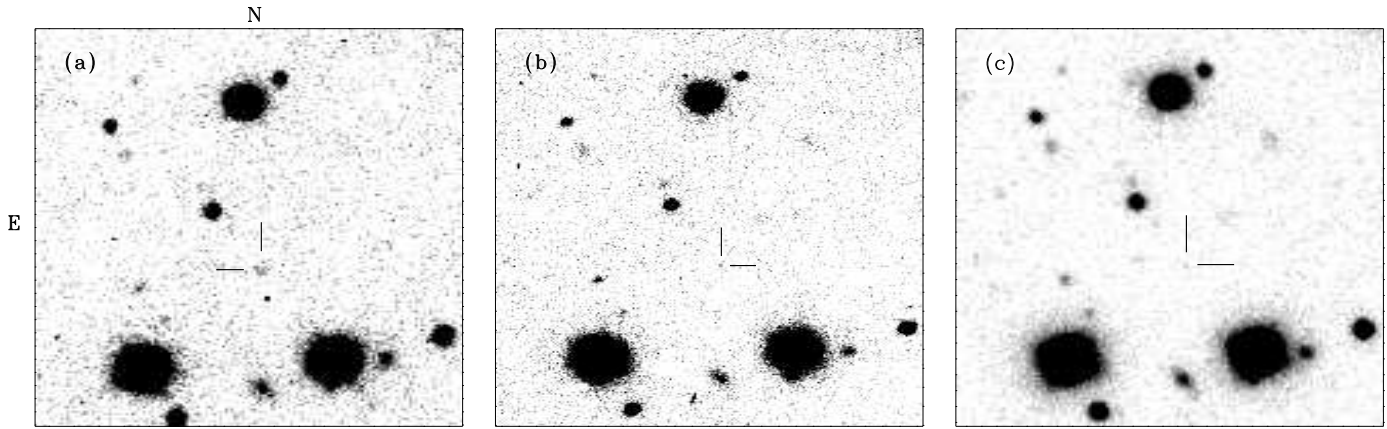,height=6cm}}
\caption{({\bf a}) NTT $R$ band image of March 29.99 UT: the indicated
object, at $R = 23.6 \pm 0.2$, is the proposed optical counterpart of
GRB980329;  ({\bf b}) sum of the $R$ band exposures taken at the NOT on
March 30.93 and 31.87; the object is still visible though fainter; ({\bf
c}) APO $R$ band image of April 1.125;  the object has faded to $R = 25.2
\pm 0.3$}
\end{figure*}

\begin{table}[t]
\begin{tabular}{l|c|c|c|c}
\multicolumn{5}{c}{Table 1. Log of Optical and Near-Infrared Observations }\\
\hline
\hline
           &          &                  &                        &          \\
Date (UT)  & Telescope&  $t^a$            & Magnitude$^b$         & Ref. \\
           &          &                  &                        &          \\
\hline
Mar 29.83 &  Teramo  &                  & $R_c>20$                  & 1  \\
Mar 29.85 & BUT, JKT &                  & $R_c>22$                  & 2  \\
Mar 29.85 &  TIRGO   &       660        & $J>19.9^c$                & 3  \\
%Mar 29.9  &   TLST  &                  & $R_c>20$                  & 4  \\
%Mar 29.9  &   TLST  &                   & $I \approx 20$            & 4  \\
Mar 29.9  &  Asiago  &                  & $R_c>21$                  & 4  \\
Mar 29.9  &  TLST    &                  & $R_c>22$                  & 5  \\
Mar 29.92 & OGS   &                     & $R_c>21$                  & 6 \\
Mar 29.99 &   NTT    &      1200        & $V>23.5$                  & 3  \\
Mar 29.99 &   NTT    &      1200        & $R_c=23.6\pm0.2$          & 3  \\
Mar 30.0  &88$^{\prime\prime}$ Hawaii & & $R_c>22.3$                & 7  \\
Mar 30.9  & BUT, JKT &                  & $R_c>22$                  & 2  \\
Mar 30.93 &   NOT    &      1200        & $R_c>25.3$                & 3  \\
Mar 30.99 &   NTT    &      1200        & $V>23.5$                  & 3  \\
Mar 30.99 &   NTT    &      1200        & $R_c>24$                  & 3  \\
Mar 31.0  &88$^{\prime\prime}$ Hawaii&  & $R_c>23$                  & 7  \\
Mar 31.5  &OAN, JKT  &                  & $R_c>22$                  & 2  \\
%Mar 31.85 &   TLST  &                   & $I >21$                   & 5  \\
Mar 31.87 &   NOT    &      1200        & $R_c>25.3$                & 3  \\
%Mar 31.92 &   NOT    &       300        & $I >19$                   & 3  \\
Apr 1.01  &   NTT    &      1200        & $V>23.5$                  & 3  \\
Apr 1.01  &   NTT    &      1200        & $R_c>24$                  & 3  \\
%Apr 1.05  &  TLST    &                  & $I >21$                   & 4  \\
Apr 1.125 &   APO    &      3600        & $R_c = 25.2 \pm 0.3$      & 3  \\
Apr 1.17  &  WIYN    &                  & $R_c>24.2$                & 8  \\
Apr 1.21  &   APO    &                  & $J>20.9$                  & 9  \\
%Apr 1.85  &  CA 2.2m &                  & $I >21$                   & 4 \\ 
Apr 1.95  &   NOT    &      1800        & $R_c>25.5$                & 3  \\
Apr 2.0   &  Keck-II    &               & $R_c = 25.7 \pm 0.3$      & 10  \\
Apr 2.0   &  Keck-I    &                & $K= 20.7 \pm 0.2$         & 11 \\
Apr 3.0   &  Keck-I    &                & $K= 20.9 \pm 0.2$         & 11 \\
Apr 3.1   &  WIYN    &                  & $R_c>23.9$                & 8 \\
Apr 6.27  &  Keck-I    &                & $K= 21.4 \pm 0.2$         & 12 \\
Apr 8.28  &  Keck-I  &                  & $K= 21.9 \pm 0.4$         & 12 \\
\hline
\multicolumn{5}{l}{$^a$ Exposure time in seconds.}\\
\multicolumn{5}{l}{$^b$ Errors are 1--$\sigma$.}\\
\multicolumn{5}{l}{$^c$ 2--$\sigma$ limit.}\\
\multicolumn{5}{l}{References: [1] Brocato et al., 1998; [2] Guarnieri et al., 
1998}\\
\multicolumn{5}{l}{[3] This paper; [4] Cappellaro, 1998; [5] Klose et
al.,
1998;}\\
\multicolumn{5}{l}{1998; [6] Corradi et al. 1998; [7] Djorgovski et al.,
1998a;}\\
\multicolumn{5}{l}{[8] Schaefer, 1998; [9] Cole et al., 1998; [10] Djorgovski et
al.,}\\
\multicolumn{5}{l}{1998b; [11] Larkin et al., 1998; [12] Metzger et al.,
1998.}\\
\end{tabular}
\end{table}

\section{Observations and Results}

\subsection{Optical}

We observed GRB980329 on March 29.99, 30.99 and April 1.01 UT at the NTT with
EMMI in $V$ and $R$ band filters, on March 30.93, 31.87, 31.92 and April 1.95 UT
at the NOT with ALFOSC in $R$ and $I$ band filters, and on April 1.125 UT at the
APO with SPICam in the $R$ band filter. The NTT observations consisted of two
10--minutes frames per filter per night, and were affected by a large airmass
(up to 3.05).  The NOT observations consisted of seven 600--seconds $R$ band
exposures and one 300--seconds $I$ band exposure.  The APO observations
consisted of six 600--seconds $R$ band exposures (see also Reichart et al.
1998).  The images have been debiased and flat-fielded following a standard
procedure. PSF-fitting photometry was done using the DAOPHOT II package (Stetson
1987) within MIDAS.  The $R$ and $V$ band images were calibrated using the
Landolt field PG1047+003 (Landolt 1992).  No calibration image is available in
the $I$ band.  We adopted airmass extinction coefficients 0.07 and 0.11 in the
$R$ and $V$ bands, respectively.  Table 1 reports a summary of all the known
optical and near-infrared calibrated observations in the first 10 days after the
GRB. 

In the sum of the two NTT $R$ band images of March 29.99 we clearly detect
an unresolved object (Fig. 1a) with $R$ = 23.6 $\pm$ 0.2 at the position
RA = 07h 02m 38s, Dec = +38$^{\circ}$ 50$^{\prime}$ 44$^{\prime\prime}$.1
(J2000), coincident (within the astrometric errors of
0$^{\prime\prime}$.4) with the transient radio source VLA J0702+3850
(Taylor et al. 1998), proposed as the radio afterglow of GRB980329, and
with the galaxy observed by Djorgovski et al. (1998a) on April 2,
tentatively identified with the host galaxy of the radio counterpart.  The
source is no longer visible in the NTT $R$ band images of the subsequent
nights, down to a limiting magnitude of $R = 24$.  Although the source is
not detected in the NOT $R$ band frames of individual nights, the sum of
the images of March 30.93 and 31.87 (corresponding to the fiducial average
UT of March 31.4, Fig. 1b)  shows an object of $R$ = 25.0 $\pm $0.5.  In
the APO $R$ band image the transient is well detected with $R$ = 25.2
$\pm$ 0.3 (Fig. 1c), indicating that the source decayed by $1.6 \pm 0.5$
magnitudes in 2.1 days.  The object is no longer detected in the NOT $R$
band exposures of April 1.95 down to a limiting magnitude of $R = 25.5$.
The source is not detected in the $V$ and $I$ bands. 

\subsection{Near-infrared}

$J$ band images of the GRB980329 field were obtained on March 29.85 at the
TIRGO with the ARNICA NICMOS 3 array detector (256x256 pixels). With the
4$^{\prime} \times$ 4$^{\prime}$ field of view detector the error box was
covered with a mosaic of frames.  The total exposure time for the central
region was 11 minutes.  Standard data reduction was performed using the
IRAF procedures in the ARNICA reduction package. 

In the final $J$ band image, calibrated using star SAO042804 (Hunt et al. 
1998), no object is detected at the position of the optical/radio
transient down to J = 20.7 (1--$\sigma$). However a 2--$\sigma$ signal at
$J$ = 20.05 is detected at RA = 07h 02m 38.13s, Dec = +38$^{\circ}$
50$^{\prime}$ 41$^{\prime\prime}$.9 (J2000), 2$^{\prime\prime}$.3 $\pm$
1$^{\prime\prime}$.0 away from the radio transient, not associated with
any other object detected in either the optical or the $K$ band.  Given
the photometric and positional uncertainties, we cannot establish the
association of this source with the GRB.

\section{Discussion}

The optical transient lies within the 1$^{\prime}$ NFI radius error box of the
X-ray afterglow. Fitting a power-law to the $R$ band flux decay, $f(t) \propto
t^{-\alpha}$, we find $\alpha = 1.3 \pm 0.2$.  This is
consistent with our non-detection of any source brighter than $R = 25.5$ in
April 1.95 and the non-detection by Guarnieri et al.  (1998).  An index $\alpha
\sim 1.3$ is also consistent with the fading observed by Klose et al. (1998) in
the $I$ band, and with our (uncertain) $J$ band measurement together with the
limit $J = 20.9$ on April 1.2 (Cole et al. 1998).  This decay is similar to that
observed in other energy ranges ($K$ band, Larkin et al. 1998; sub-mm, Smith \&
Tilanus 1998; X-rays, in 't Zand et al. 1998) and to the decays of previous
optical afterglows of GRBs ($\alpha \simeq 1 \div 2$, van Paradijs et al. 1997,
Fruchter et al. 1998, Galama et al. 1998, Diercks et al. 1998, Groot et al. 
1998). We conclude that the transient object is the fading counterpart of the
GRB980329.  Extrapolating the power-law to the time of the Keck observations by
Djorgovski et al. (1998b)  yields a magnitude consistent with the one they
report for the putative host galaxy ($R$ = 25.7 $\pm$ 0.3).  This suggests that
the optical transient could still contribute a significant fraction of the total
observed flux at that time. 

\begin{figure}[t] 
\psfig{figure=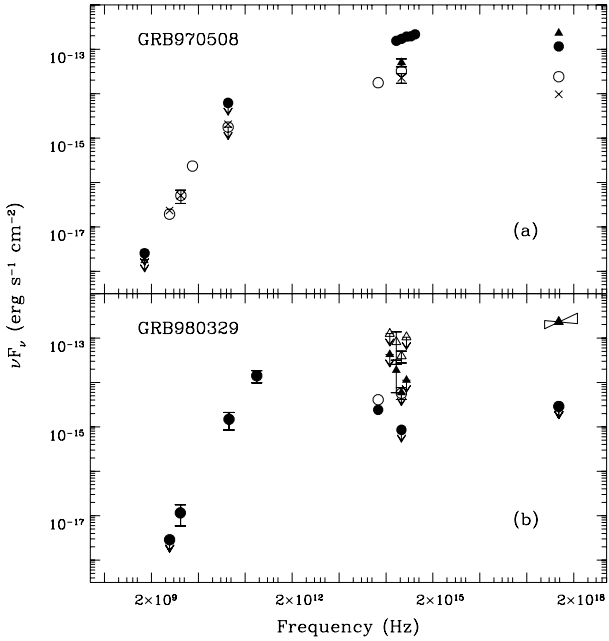,width=9cm} 
\caption{Comparison of multi-wavelength energy distributions of GRBs
970508 and 980329 corrected for the Galactic extinction (see text). ({\bf
a}) Energy spectrum of GRB970508 in May 9.9 (filled triangles); May 10.8
(filled circles); May 16.9 (open circles); May 19 (crosses).  Data are
from Piro et al. (1998, X-ray); Galama et al. (1998, $U$ photometry); 
Sokolov et al. (1998, $BVRI$); Castro-Tirado et al. (1998, $R$); Pedersen
et al.  (1998, $R$); Morris et al.  (1997, $K$); Bremer et al. (1998, mm); 
Frail et al. (1997, radio).  Error bars have been omitted when smaller
than the symbol size.  ({\bf b})  Same as (a) for GRB980329 in March 29.99
(filled triangles) and April 6.2 (filled circles). Data are taken from in
't Zand et al. (1998, X-ray; the slope of the BeppoSAX MECS spectrum is
reported along with its 1--$\sigma$ confidence range);  
Smith \& Tilanus (1998, mm); Taylor et al. (1998, radio). 
See Table 1 for references to the optical and near-infrared photometry. 
Data corrected for typical intrinsic obscuration within starburst galaxies
at $z = 1$ are also reported as open triangles for March 29.99 and open
circles for April 6.2}
\end{figure}

Since GRB980329 and GRB970508 are the only two GRBs so far with radio and
mm counterparts we have compared their radio-to-X-ray spectral energy
distributions (SEDs, $\nu$$f_\nu$) with the multi-wavelength afterglow
model of Sari et al. (1998).  In Figures 2a and 2b are reported the SEDs
of GRB970508 and 980329, respectively, at different epochs.  For GRB970508
they correspond to the start of rise to optical maximum (May 9.9); to the
optical maximum (May 10.8); to the last upper limit determination in the
sub-mm band (May 16.9); to the first sub-mm detection (May 19).  For
GRB980329, the chosen epochs correspond to the first $R$ band detection
(March 29.99) and to the second sub-mm detection (April 6.2).  The data
(see figure caption for references) have been corrected for the
interstellar extinction within the Galaxy, $A_V = 0.09$ for GRB970508
(Djorgovski et al.  1997), $A_V = 0.4$ for GRB980329 (Rowan-Robinson et
al. 1991).  We adopted the Galactic extinction curve of Cardelli et al. 
(1989). The X-ray data of GRB970508 on May 16.9 and 19 are extrapolated in
time according to the power-law used to model the X-ray light curve (Piro
et al. 1998). For the SED of GRB980329, the point at 8.46 GHz at the
second epoch is the average of the three measurements of Taylor et al. 
(1998). We made a rough calibration of the March 29.9 $I$ band data (Klose
et al., 1998) by a power-law extrapolation of the $R$ and $V$ band data,
inferred from the March 29.99 NTT images, of a nearby object that appears
to be equally bright as the transient in the $I$ band image.  From this we
infer that, if the nearby object remained constant, the GRB should have
$I\simeq 21.5$ on March 29.9. We have scaled this value to March 29.99,
according to a power-law of index 1.3 and assigned a conservative error
bar of $\sim0.7$ magnitudes, to account for the uncertainties in spectral
slope, temporal extrapolation and eye estimate of the relative $I$ band
fluxes of the two objects in Klose et al.'s images. For the second epoch,
we used the magnitude of the host galaxy, $R = 25.7$, as an upper limit on
the level of the transient.  In fact, all optical and near-infrared
magnitudes of the point-like source might be overestimated, due to the
underlying host galaxy, the brightness of which is not precisely
determined in any band. The X-ray flux at 4 keV on March 29.99 has been
computed using a temporal power-law of index $\alpha = 1.03 \pm 0.13$ and
a spectral index $\beta = 0.8 \pm 0.3$ (in 't Zand et al. 1998; $f_\nu
\propto \nu^{-\beta}$).  A temporal decay with $\alpha > 1.8$ (Greiner et
al. 1998) has been assumed to compute the upper limit on April 6.2. 

The $I$ band detection of Klose et al. (1998), roughly calibrated by us ($I
\sim 21.7$), and our $R$ band measurement on March 29.99 (see Table 1)  yield
a color $R - I \sim 2$, equivalent to a spectral index of $\sim 5$, which
suggests a red spectrum for the transient. Similarly, from our $R$ and $J$
band quasi-simultaneous measurements we derive an upper limit of $R - J ~
\mincir 3.6$, consistent with a red spectrum ($\beta ~ \mincir 4$).  The
measurements in April 2 (Table 1) give $R - K ~ \magcir 5$, corresponding to
$\beta ~ \magcir 3$.  These spectral slopes are much steeper than found
for GRB970508, for which spectroscopy (Metzger et al. 1997)  and multi-band
photometry (see caption to Fig. 2 for references) give an optical index of
$\beta \simeq 1$ and an optical-to-near-infrared index of $\beta
\simeq 0.5$. Moreover, the blast wave models of GRB afterglows predict flatter
indices in these bands ($\beta \simeq 0.5 \div 1.5$, see Sari et al. 
1998, Wijers \& Galama 1998). Since the dust column density in the direction
of GRB980329 is moderate, reddening could be due either to intrinsic or
intergalactic extinction.  The former would imply the presence of a dust and
gas rich medium and likely intense star formation, while the latter would
suggest a rather high redshift. The remarkable intensity of the burst does not
point to a very large distance (see also in 't Zand et al. 1998), therefore we
have considered the possibility that GRB980329 occurred in a starburst galaxy,
and we have investigated how the typical obscuration of this class of galaxies
would affect the intrinsic spectrum of GRB980329 at a range of redshifts up to
$z \sim 2$.  To this aim, we corrected the SED of GRB980329 for the redshifted
extinction curve of 19 local starburst galaxies (Calzetti 1997).  Since the
curve has been derived by considering only the starburst regions, excluding
the old stellar populations of the sample galaxies, we expect it to be
appropriate also for starburst galaxies at moderate or high redshift (Calzetti
1998).  We have noted that for $z \simeq 1$ the corrected $R$ and $I$ band
flux ranges of March 29.99 are consistent with the extrapolation to optical
frequencies of the power-law which best fits the X-ray spectrum, and that the
corrected $R$ band upper limit and $K$ band flux of April 6.2 imply a spectral
index $\beta \ge 0.7$, consistent with theoretical expectation (see Fig.
2b).  This redshift value is in agreement with the observation of intense star
formation at $z \sim 1$ (Madau et al. 1996) and thus broadly supports the
proposal that GRBs occur in actively star-forming galaxies (Paczy\'nski 1998).
The visual extinction of the applied model ($A_V = 1.2$) would correspond to
$N_H = 1.9 \times 10^{21}$ cm$^{-2}$ in our Galaxy (and therefore to a
probably not higher value in a starburst), well within the range of $N_H$
derived from the power-law fit to the X-ray data (in 't Zand et al. 1998). 

The multi-wavelength data of the two afterglows are too few and sparse
to allow a detailed description of the spectral shapes in the various
bands and precise localization of the energy peaks.  However, the
overall SEDs of the two bursts at the different epochs seem to be
consistent with a single emission component, whose maximum shifts in
time toward lower energies.  For GRB970508, the SED appears smooth and
exhibits a broad peak which, at the first epoch, likely falls at or
immediately above X-ray frequencies, while it is located between
$10^{14}$ and $10^{16}$ Hz on May 19. In GRB980329, the peak of the
$\nu$$f_\nu$ curve on March 29.99 is at or around $10^{18}$ Hz, as
indicated by the X-ray spectral index, while on April 6.2 it is located
between $10^{12}$ and $10^{14}$ Hz and appears narrower.  According to
the Sari et al.  model, in which the emission is produced through
synchrotron radiation from an expanding relativistic shell, it is
possible to reproduce the observed temporal evolution of the
multi-wavelength spectrum of GRB970508 for a relativistic electron
distribution $N(\gamma) \propto \gamma^{-p}d\gamma$ with $p ~\magcir 2$. 
(Notice that the representation of the synthetic spectrum in their Fig.
1 is in $f_\nu$.)  On the other hand, the SED peak of GRB980329, which
is much better constrained than that of GRB970508, exhibits a
displacement by more than 4 decades in energy in $\sim$10 days.  This
cannot be accounted for by the Sari et al. model, even assuming the
fastest temporal evolution envisaged by their scenario (fully radiative
expansion in fast cooling regime, which holds anyway only within short
times after the GRB).  The onset of a subrelativistic expansion some
days after the GRB would imply a faster change of the synchrotron break
frequency of the electron distribution (Wijers et al. 1997) and might
better describe the observed behaviour.

The hypothesis of significant intrinsic obscuration at $z \sim 1$ for
GRB980329 improves the modeling of its multi-wavelength data, although it
does not account for its apparently more abrupt spectral evolution than
observed for GRB970508.  Under this assumption, which by no means
represents a stringent prediction on the redshift, we estimate that the
total energy released by the GRB in the observed 40-700 keV range would be
approximately $3 \times 10^{52}$ erg ($H_0$ = 65 km s$^{-1}$ Mpc$^{-1}$,
$q_0 = 0.5$), consistent with models of GRB formation from coalescence of
neutron stars. 

\begin{acknowledgements} 
We thank G. Tozzi for assisting with data acquisition, D. Calzetti, J. 
Gorosabel, and L. Hunt for useful discussion, and M. Salvati for critical
comments to the manuscript.

\end{acknowledgements}

\end{document}